# Optical bistability in a one-dimensional photonic crystal resonator using a reverse-biased pn-junction

Majid Sodagar, Mehdi Miri†, Ali A. Eftekhar, and Ali Adibi*

¹School of Electrical and Computer Engineering, Georgia Institute of Technology, 777 Atlantic Drive NW, Atlanta, GA 30332, USA
†Current address: School of Electrical and Computer Engineering, Shiraz University, Engineering Faculty No. 1, Zand St., Shiraz, Iran
*ali.adibi@ece.gatech.edu

**Abstract:** Optical bistability provides a simple way to control light with light. We demonstrate low-power thermo-optical bistability caused by the Joule heating mechanism in a one-dimensional photonic crystal (PC) nanobeam resonator with a moderate quality factor ($Q \sim 8900$) with an embedded reverse-biased pn-junction. We show that the photocurrent induced by the linear absorption in this compact resonator considerably reduces the threshold optical power. The proposed approach substantially relaxes the requirements on the input optical power for achieving optical bistability and provides a reliable way to stabilize the bistable features of the device.

## 1. Introduction

Use of nonlinear properties of optical materials is inevitable for all-optical signal processing applications [1]. In general nonlinear optical coefficients such as the third order nonlinear susceptibility ($\chi^3$), in common integrated photonic materials, i.e. silicon (Si) and silicon nitride (SiN), are relatively weak and can be neglected in low field intensity limits [2,3]. Nevertheless, in high-$Q$ microresonators with small effective mode volumes, the field intensity (and optical power density) can reach to a rather high level such that nonlinear effects become significant even at modest input powers.

 Optical bistability in passive optical resonators, as one manifestation of nonlinear effects, has long been of interest as a simple mechanism to manipulate light with light. This phenomenon has been observed and carefully studied in various material platforms and different types of integrated resonators including one/two-dimensional PC resonators as well as microdisk and microring resonators [4-12]. It has been shown that the interplay between the heat generation due to optical absorption at resonance wavelength and the shift in the resonance wavelength of the resonator caused by the thermo-optic effect of the resonator material can result in a hysteretic behavior and a positive feedback process which can lead to optical bistability. The thermal bistability is especially important in Si photonics due to the widespread use of Si on the device layer for integrated photonic structures. The thermal bistability in most of the reported Si photonic devices to-date rely on the nonlinear optical absorption processes, i.e., two-photon absorption (TPA) and the subsequent free-carrier absorption (FCA). Thermal-bistability can also be achieved through linear absorption in optical resonators [13]. However, in both cases of thermal bistability, the threshold field density in the resonator is very high. This severely limits the use of optical bistability to form practical nonlinear integrated photonic devices in Si.

 In this paper, we demonstrate low-power thermal-bistability in Si photonic resonators with moderate Qs (e.g., 10 K) based on amplification of the heat generated by linear absorption using Joule-heating [14]. In our designed device, a reverse-biased pn-junction is integrated with a nanobeam PC resonator. The illumination of the depletion region of the pn-junction by the incident light results in photo-generated carriers through linear absorption in Si, and these carriers are swept by the electric field (due to the reverse bias) and are collected at the two sides of the pn-junction device [15-18]. Such photocurrent, though small, once accompanied by a large reverse bias can considerably increase the generated heat per absorbed photon through Joule-heating to considerably reduce the optical power threshold for achieving optical bistability. Our experimental studies show that the photocarriers are induced mainly by the linear optical absorption in the device. The proposed approach can be exploited to design ultralow power optical transistors, optical memory elements, and ultrasensitive sensors [19-22]. In addition to its contribution to heat generation, the photocurrent can be monitored in an external electronic circuit to generate a negative feedback signal (i.e., adjusting the reverse bias voltage) to stabilize the bistable features against the random variations of the environmental conditions (e.g., temperature).

## 2. Nanobeam resonator design

The PC nanobeam resonator used in this study is designed and fabricated in a silicon-on-insulator (SOI) substrate with a 250 nm thick Si layer on a 3 μm buried oxide layer. The resonator comprises a symmetric resonant region sandwiched between two symmetric PC mirrors. The resonant region and the PC mirrors consist of a number of air holes with their centers located in a one-dimensional (1D) period lattice with lattice constant $a$ (see Fig. 1(a)). These air holes are etched in a Si rib waveguides (WG) with a width of $w$ sitting on a 50 nm thick pedestal. The radii of the air holes in the mirror regions are equal to $r_m$ while the radii of the air holes in the resonant region are different from each other, and they are optimally engineered to maximize the quality factor. The overall structure is symmetric around its center. The resonant region has 15 pairs of air holes, and each mirror section contains 5 air holes. The number of air holes for the two mirrors is chosen to ensure a decent coupling $Q$ ($Q_w$) between the resonator and the inline WG at the two terminals of the device.

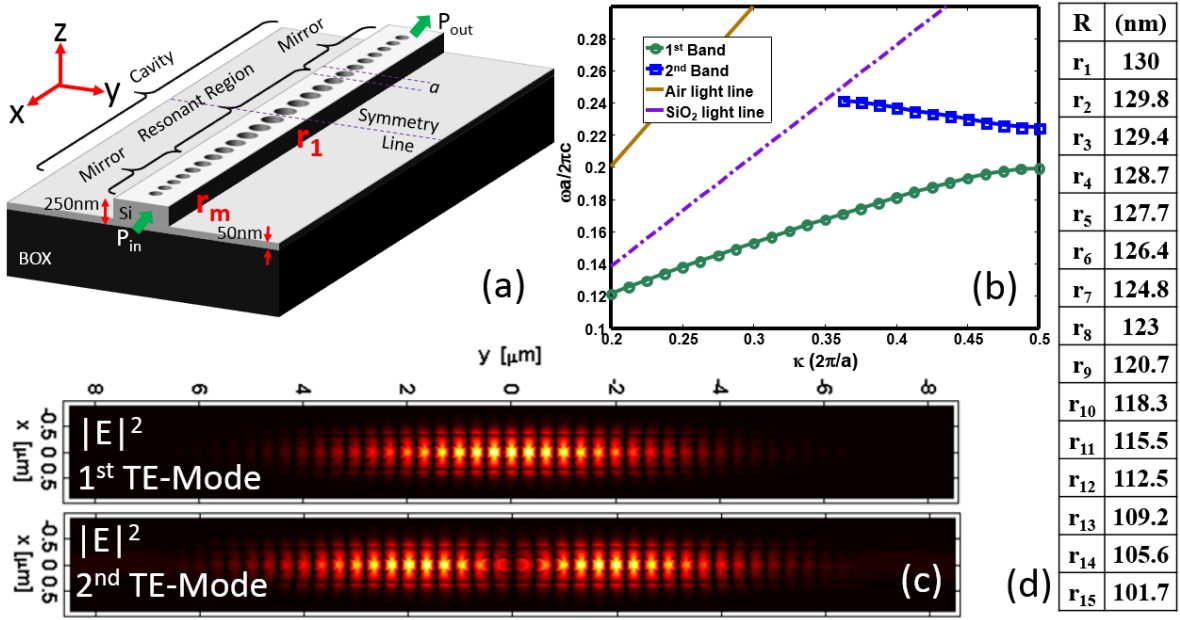

Fig. 1. (a) The 3D schematic of the nanobeam PC resonator; (b) Normalized band diagram of the periodic mirror regions showing a photonic band gap in the range 181 THz < $f$ < 204 THz (for $a$ = 330 nm, the corresponding wavelength range is 1469 nm < $\lambda$ < 1656 nm ); (c) The field profiles of the first ($\lambda_1$=1579.71 nm) and the second ($\lambda_2$=1609.95 nm) TE resonant modes of the device in (a) with mode volumes of $0.97(\lambda_1/n_{si})^3$ and $1.36(\lambda_2/n_{si})^3$ respectively;(d) Tabulated air-hole radii calculated via the TL technique for the resonant region in part (a).

Given the geometrical dimensions of the substrate, the width of the WG is chosen at $w$ = 700 nm to support a propagating mode in the desired wavelength range (i.e., 1469 nm < $\lambda_0$ < 1656 nm) well below the light lines. The lattice constant is set to be $a$ = 330 nm. This choice will place the edge of the Brillouin zone (i.e., $\kappa = \pi/a$, with $\kappa$ being the normalized propagation constant) far from the edge of the radiation zone (i.e., $2\pi/\lambda_0$) to reduce the radiation loss [23]. The photonic bandgap of the PC in the mirror section is adjusted by setting the air hole radius to $r_m = 100$ nm to place the desired resonance wavelength at the center of the photonic gap (see Fig. 1(b)). In order to maximize the radiation $Q$ ($Q_r$) of the resonant region (i.e., minimize the radiation loss), we use the transmission line (TL) technique (detailed in Ref. [24]) to design the air holes in the resonant region such that the spatial profile of the resonant mode (see Fig. 1(c)) meets all the requirements for low radiation loss explained in Ref. [25]. The calculated air hole radii for the resonant region are tabulated in Fig. 1(d).

We use a commercially available finite difference time domain (FDTD) based software (i.e., Lumerical) to obtain the actual mode profiles as well as estimates of the optical $Q$s of the resonant modes of the resonator device in Fig 1(a). Figure 1(c) shows the magnitude of the electric field for the first and the second TE (electric field in the XY plane in Fig.1 (a)) mode at $\lambda_1 = 1579.71$ nm, $\lambda_2 = 1609.95$ nm (Si refractive $n_{si} = 3.46$). To separately estimate the $Q_r$ and $Q_w$ of these modes, we simulate the structure with (a) 15 and (b) 5 PC lattice periods in the mirror regions. Simulations suggest that the leakage power from the fifteen-period mirrors is negligible (i.e., $Q_w \rightarrow \infty$), and the estimated $Q$s of $Q_1=Q_{r1} \approx 1.36 \times 10^6$ and $Q_2=Q_{r2} \approx 3.77 \times 10^5$ for the first and the second modes, respectively, are mainly limited by the radiation loss (note that we neglected the material loss due to its negligible effect on Q at the selected wavelengths). Also, $Q$s of $Q_1=Q_{w1} \approx 48000$ and $Q_2=Q_{w2} \approx 3700$ (i.e., $Q$s are limited only by the coupling loss to the input and output WGs) are estimated for the first and the second modes, respectively, when five-period mirrors are used. By comparing the values of $Q_r$ and $Q_w$ obtained for each mode, we conclude that the Q of each mode under a practical design for achieving reasonable coupling to the input/output WGs will not be limited by the radiation Q, and they will be defined by the coupling Q (i.e., $Q_w$).

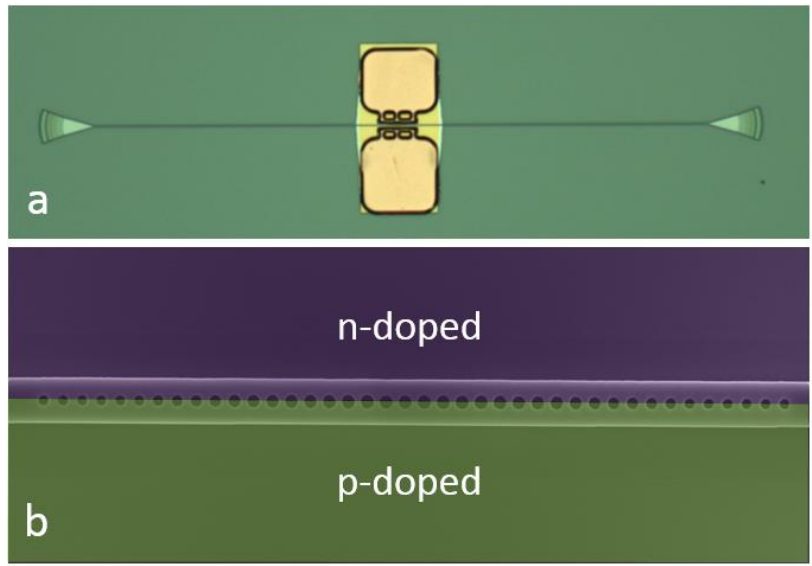

Fig. 2. (a) Optical micrograph of the fabricated device showing the copper pads on a 50 nm thick pedestal around the PC nanobeam resonator as well as the 400 μm long inline feeding WG along with the focusing grating couplers at the two ends. (b) False-colored SEM of the fabricated PC nanobeam resonator (taken before metallization). The purple and green colors represent the n-type and p-type regions, respectively.

To complete the design of the structure, we add two focusing grating couplers on the input and output rib waveguides for out-of-plane coupling of the input and output of the overall device to the input laser and output detector, respectively [26]. These gratings are designed using the conventional techniques and for the sole purpose of characterizing the device.

## 3. Fabrication

The PC nanobeam resonator along with the input/output focusing grating couplers (designed to have a peak coupling efficiency at around 1600 nm) are fabricated on an SOI wafer with a 250 nm thick Si layer. A 110 nm layer of 6% hydrogen silsesquioxane (HSQ from Dow Corning) is spin coated on the sample as the resist, and the patterns are defined through electron beam lithography (EBL). After development in 25% tetramethylammonium hydroxide (TMAH) at an elevated temperature of 40 ºC, the device Si layer is etched with an etch-depth of 200 nm in an inductively-coupled-plasma (ICP) chamber. This step leaves a 50 nm pedestal on the sample, which is removed selectively around the grating coupler areas. The sample is then covered with 10 nm of silicon oxide ($SiO_2$) through atomic layer deposition. The metallurgical pn-junction in the middle of the nanobeam structure is realized by successive ion implantation steps to achieve concentrations of ~ $6\times10^{17}$ $cm^{-3}$ on the resonator region (75As+ and 49BF$_2$+ species) and ~ $10^{20}$ $cm^{-3}$ on the contact regions. This level of doping increases the material loss in the Si layer to $\alpha = 3.35$ $cm^{-1}$ at the telecommunication wavelength [27]. The drop in $Q$ of the designed nanobeam resonator due to additional material loss is estimated through the perturbation theory using the obtained mode profiles shown in Fig. 1(c) [28], and the calculated values are ~ 9000 and ~ 3500 for the first and the second TE modes, respectively. The patterned polymethyl methacrylate (PMMA from MicroChem) resist (thickness ~ 2 μm) is used as the implantation mask in all doping steps. In the next step, dopants are electrically activated by annealing the sample at 950 ºC for 240 seconds in a rapid thermal processing (SSI RTP) system. For the metallization layer, titanium (as an adhesion layer) and copper are sputtered successively on another patterned layer of PMMA resist (thickness ~ 3 μm) and lifted off with the aid of a sonication bath. The optical micrograph of the device along with the scanning electron micrograph (SEM) of the resonator region is shown in Figs. 2(a) and 2(b), respectively. The nanobeam is connected to the feeding WG at both ends of the resonator. The focusing gratings at the two ends of the feeding WGs facilitate the input/output coupling of light to/from the device. The optical field propagating inside the WG couples into and out of the PC nanobeam resonator through the mirror regions at the two ends of the resonator.

**4. Characterization**

To characterize the device, the output light of a cw tunable laser (Agilent 81682A) is launched into a cleaved single mode fiber through an in-line polarization controller. The fiber is mounted on a stage equipped with XYZ micro-positioners as well as a rotation/tilt compensator. The fiber is then aligned so that its outcoming light is focused on the input grating coupler. Similarly, the output light from the chip is collected through the output grating coupler with a similar cleaved fiber and fed directly into a detector (Thorlabs PDB150C 800-1700 nm). The chip is placed on a temperature controlled stage with the temperature set at 25 ºC.

Figure 3 shows the normalized transmission spectrum of the device, which is obtained by sweeping the laser wavelength from 1510 nm to 1640 nm at the rate of 5 nm per second. To have a good signal-to-noise ratio, the output power of the laser is set to its maximum-over-sweep value, i.e., 354 μW. We also repeat the measurement with the laser power set at a much lower level of 5 μW. Since the later measurement results in the same lineshape for both resonances, we conclude that the nonlinear loss sources have negligible effect on the measurement with the higher laser power, and all observed effects are due to linear phenomena. For normalization, we divide the transmission spectrum to that of a reference device (fabricated on the same chip) that consists of similar grating couplers and WGs with no PC nanobeam resonator in the middle (same overall device length).

The two pronounced peaks in the collected spectrum in Fig. 3 are related to the two supported resonant modes of the PC nanobeam resonator. The first TE mode at 1598.8 nm features a loaded-$Q$ ($Q_L$) of ~ 8900. The second TE mode at 1627.3 nm is closer to the bandgap edge and exhibits a larger spatial extend along the resonator and consequently experiences a stronger coupling from and to the feeding WGs (Fig. 1). The measured $Q_L$ for this mode is ~ 3200. The actual resonance wavelengths are slightly higher than the simulated ones owing to the thin deposited silica on the sample.

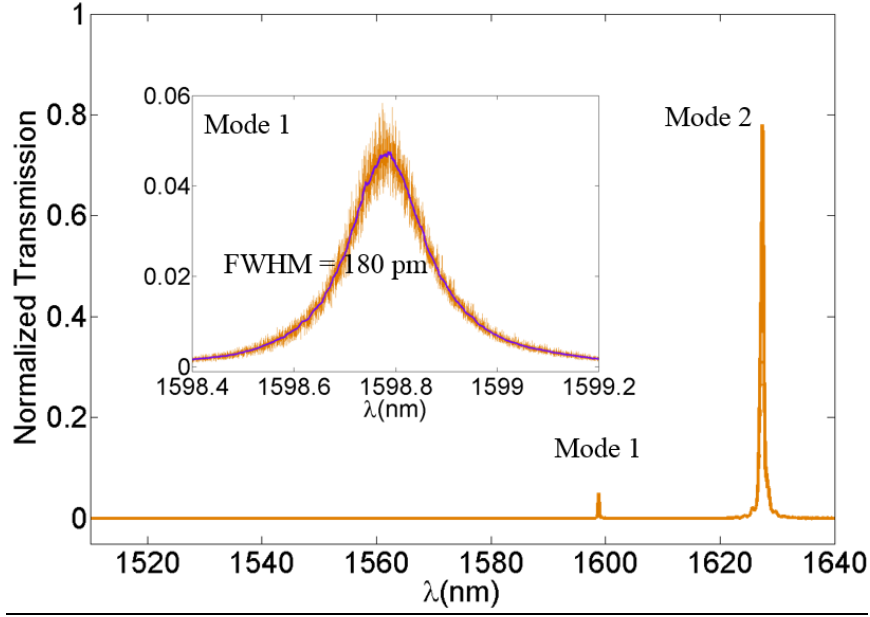

Fig. 3. Measured normalized transmission spectrum of the PC nanobeam resonator in Fig. 2. Inset shows a closer look at the linewidth around the first mode. The designed photonic bandgap of the PC mirrors covers the range 1469 nm < λ < 1656 nm.

The measured transmission of the resonator at the first and second resonance wavelengths are 0.047 and 0.76, respectively. With the symmetric resonator-to-input/output WG coupling regions, the resonator transmission at resonance is given by $T(\omega_0) = (Q_L/Q_w)^2$ [28]. From this relation, the experimental $Q_w$ values can readily be calculated to be 41000 and 3670 for the first and the second modes, respectively which are in good agreement with the predicted values from FDTD simulations. We use a calibrated optical head power sensor (Agilent 81624A) to measure the actual power coupled into the input fiber and the power collected at the output fiber in the reference device. From such measurements, the sum of the coupling loss of the grating coupler and the propagation loss in the WG region is estimated to be ξ ~ 6.6 dB at the first resonant wavelength (i.e., 1598.8 nm). The $E_c = (Q_w/\omega_0)P_{out}$ expression (see Ref. [28]) is then used to estimate the total stored electromagnetic energy in the resonator ($E_c$) for different laser output powers. In this expression $P_{out}$ is the WG output power (see Fig. 1(a)) which is estimated by measuring the power in the output fiber of the actual device and normalizing it to the propagation losses (i.e., ξ). Plots in Fig. 4 (a) show the transmission spectrum of the device around the first resonant mode for different laser excitation powers with no reverse bias applied to the pn-junction (the corresponding resonator energies are provided in the legend). It is seen that for the applied laser power levels, the resonance wavelength and the lineshape are preserved. This observation suggests that the nonlinear absorption mechanisms are not present at zero reverse-bias.

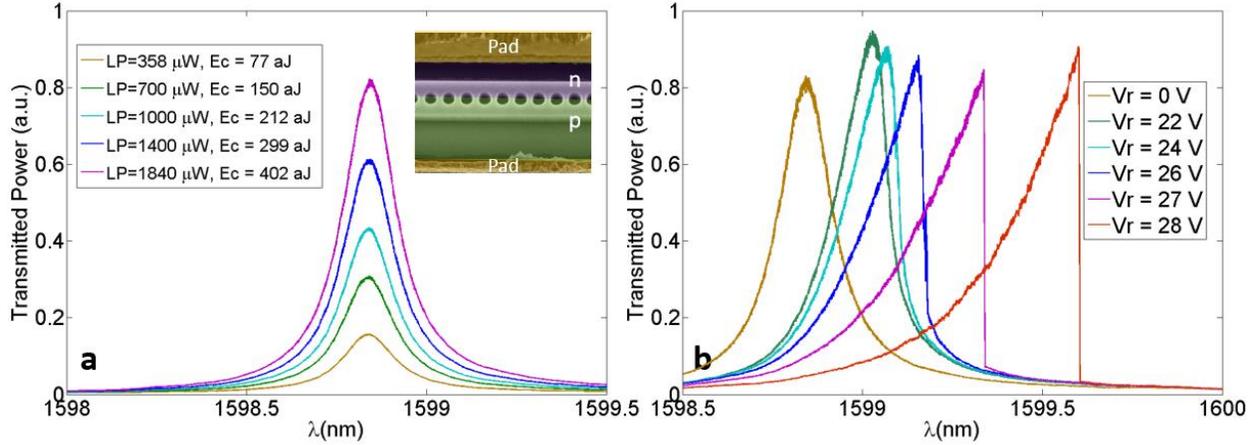

Fig. 4. (a) Measured transmitted power spectrum of the nanobeam resonator at different laser output powers ($P_L$) with the applied voltage of the pn-junction device kept fixed ($V_r = 0$), (b) Transmitted power spectrum of the nanobeam resonator at a fixed laser power ($P_L = 1.84$ mW) with varying reverse bias applied to the pn-junction device.

To observe the effect of the applied reverse bias of the pn-junction device, we monitor the transmission spectrum of the device under different reverse-bias voltages, and the results are shown in Fig. 4(b). In this measurement, the laser wavelength is swept from shorter to longer wavelengths. As shown in Fig. 4(b), the Lorentzian lineshape of the resonance is no longer preserved for bias voltages larger than 22 V. The observed redshift in the resonance wavelength for smaller voltages (i.e., $V_r < 22$ V) is attributed to the carrier dispersion property of Si, which is triggered by the depletion of the majority carriers in the p and n regions [27]. Once the applied bias voltage exceeds 22 V, the resonance linewidth becomes lopsided, and the resonance peak experiences an even bigger redshift; this is not explicable only through the carrier dispersion property of Si. As seen in Fig. 4(b), the resonance lineshape in this regime broadens and features an abrupt jump on the right side (higher wavelength) of the peak. This behavior is known as the peak dragging which is directly linked to the bistability condition [29]. This characteristic can be explained by considering both the thermo-optic effect in Si ($dn_{si}/dT = 1.86 \times 10^{-4}$ $K^{-1}$ [30]) and the Joule heating mechanism due to the induced photocurrent in the device. Once the laser wavelength reaches to the vicinity of the resonance wavelength (from the left side) the optical field in the resonator generates a finite photocurrent. The associated dissipated electric power increases the device temperature which in turn results in a small redshift in the resonance wavelength. This redshift in the resonance wavelength tends to reduce the optical field in the resonator and hence, it limits the photocurrent generation. This negative feedback process turns into a positive one once the thermo-optic effect can no longer catch up with the Joule heating effect as the laser wavelength is swept to longer wavelengths. The abrupt jump in the spectrum is where this condition manifests itself.

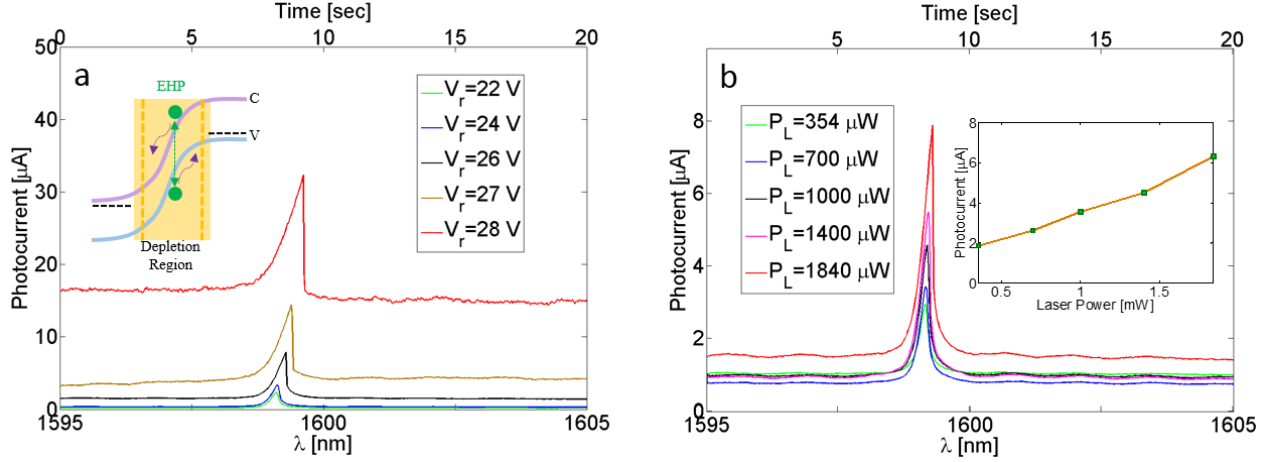

Fig. 5. (a) Measured pn-junction leakage current in the resonator region as the laser wavelength is swept (the top and bottom horizontal axes are the sweeping time and the corresponding laser wavelength, respectively). In these measurements the laser power is kept fixed at $P_L$= 1.84 mW, (b) Measured photocurrent generated for different laser power (bias is kept fixed at $V_r$= 22 V) as a function of sweeping time. Inset shows the photocurrent jump versus the laser power ($P_L$).

We use a source measurement unit (SMU) (Keithley 4200-SCS) to measure the photocurrent generated in the pn-junction. In this experiment, we slowly sweep the wavelength of the tunable laser (at a rate of 500 pm/s) from 1595 nm to 1605 nm. This range covers the first resonance wavelength of the nanobeam resonator. Different curves in Fig. 5(a) show the measured leakage current of the device at different applied reverse-bias voltages as a function of the laser scanning wavelength. In this study, the output power of the laser is fixed at 1.84 mW (corresponding to 403 µW in the input waveguide). As it can be clearly seen, the leakage current rises at wavelengths close to the resonance wavelength of the resonator because of the increased field density in the resonator. Application of higher reverse bias voltages widens the depletion region, which modestly increases the photocarrier collection efficiency. Also the generated photocarriers experience a stronger driving force under higher electric fields. It is seen in Fig. 5(a) that the generated photocurrent progressively increases from ~ 1.9 µA to ~ 16 µA as the reverse-bias voltage increases from 22 V to 28 V. The leakage current (off-resonance) is also rises with the reverse-bias voltage. The plots in Fig. 5(b) show that the photocurrent increases in response to the successive increments in the laser power at a fixed reverse-bias voltage, i.e., $V_r$ = 22 V. It is seen that by increasing the laser power from 354 µW to 1.84 mW, the photocurrent rises from ~ 1.9 µA to ~ 6.4 µA rather linearly (see the inset in Fig. 5(b)). This linear characteristic suggests that the photocarriers are generated mainly through the linear absorption mechanisms.

## 5. Discussion

Although pure bulk Si has minimal linear absorption at telecommunication wavelengths, additional implantation, material defects, and also the mid gap surface states, e.g., at the Si-air or Si-$SiO_2$ interfaces, substantially contribute to the linear absorption of Si-based nanostructures [30-35]. Following the expressions in Appendix 1, we can estimate the linear and the nonlinear absorption rates. Assuming a perfect collection of photocarriers, i.e., $\eta = 1$, the collective linear absorption rate is calculated to be $\gamma_L = 2.9 \times 10^{10}$ s$^{-1}$ when the stored energy is $E_c = 402$ aJ (corresponding to 397 µW input waveguide power and the resonator Q of 8900). The TPA and FCA absorption rates are also evaluated to be $\gamma_{TPA} = 5.50 \times 10^7$ s$^{-1}$ and $\gamma_{FCA} = 2.46 \times 10^7$ s$^{-1}$ for the same stored energy.

It is seen that the nonlinear absorption rates are much smaller than the collective linear absorption rate. This result is consistent with the experimental observations detailed in the characterization section. This confirms that the generated photocurrent in our experiments is mainly due to the linear absorption processes. It is worth noting that in principle; the photocarriers due to TPA can also (before non-radiative relaxation takes place) contribute to the photocurrent. Such photocarriers can be generated in high-$Q$/low-mode volumes cavities at low input optical powers as shown in numerous works [7,8]. In such cases optical bistability can be achieved even at lower input optical powers once the structure is integrated with a similar revered-biased pn-junction explained in the present work.

## 6. Conclusion

We demonstrated here an optical bistable device in a Si-based integrated photonic platform comprising a rather low-$Q$ PC nanobeam resonator with an embedded pn-junction. In contrast to other Si-based bistable optical devices, which depend on heating through non-radiative relaxation of carriers (triggered by nonlinear TPA and FCA processes), the bistable functionality of the proposed device relies on the interplay between the generated heat due to the photocurrent and the thermo-optic effect. We demonstrated that the generated photocurrent through linear absorption mechanisms is sufficient to achieve bistability. The proposed Joule-heating mechanism is particularly of interest as it permits the realization of optical bistability at comparatively low input optical powers. The required input optical power in the device used in this work can be further reduced significantly by improving the photocarrier collection efficiency and using a resonator with higher $Q$. The proposed mechanism allows for design and implementation of all-optical processing systems and robust ultrahigh sensitive sensors at extremely low optical powers.

## Acknowledgment

This work was supported by the Air Force Office of Scientific Research under Grant No. FA9550-13-1-0032 (G. Pomrenke). We would also like to thank Dr. Vogel's group at Georgia Tech for sharing their measurement equipment with us.

## Appendix 1

We define $\gamma_L$ as the collective linear absorption rate which contributes to photocarrier generation process. $\gamma_L$ can readily be linked to the generated photocurrent ($I_p$) by the following equation:

$$I_p = \eta \frac{q\lambda}{hc} \gamma_L E_c. \tag{A.1}$$

In Eq. (A.1), $q$ represents the electron charge, $\eta$ is the carrier collection efficiency, $\lambda$ is the resonance wavelength in vacuum, $c$ is the speed of light, and $h$ is Planck's constant.

The nonlinear absorption rates depend on the resonator mode profile ($E(r)$) as well as the stored energy, i.e., $E_c$. The TPA rate ($\gamma_{TPA}$) can be estimated through the following expressions [5]:

$$\gamma_{TPA}(E_c) = \Gamma_{TPA} \frac{\beta_{Si} c^2}{n_{Si}^2 V_{TPA}} E_c, \quad V_{TPA} = \frac{\left(\int n^2(r) E^2(r) dr\right)^2}{\int n^4(r) E^4(r) dr}, \quad \Gamma_{TPA} = \frac{\int_{Si} n^4(r) E^4(r) dr}{\int n^4(r) E^4(r) dr}. \tag{A.2}$$

In Eq. (A.2), $\beta_{Si} = 8.4 \times 10^{-12}$ mW$^{-1}$ is the TPA coefficient of Si [2]. In Si with a reverse-biased pn-junction the effect of intrinsic free-carriers is minimal. However the FCA due to the free carriers induced by TPA can be significant. Similarly we have the following expressions for FCA rate ($\gamma_{FCA}$) [5].

$$\gamma_{FCA}(E_c) = \Gamma_{FCA} \left( \frac{\tau \sigma_{Si} c^3 \beta_{Si}}{2 n_{Si}^3 \hbar \omega_0} \frac{E_c^2}{V_{FCA}^2} \right), \quad V_{FCA}^2 = \frac{\left(\int n^2(r) E^2(r) dr\right)^3}{\int n^6(r) E^6(r) dr}, \quad \Gamma_{FCA} = \frac{\int_{Si} n^6(r) E^6(r) dr}{\int n^6(r) E^6(r) dr}. \tag{A.3}$$

In Eq. (A.3) $\tau$ represents the free-carrier lifetime, which strongly depends on the free-carrier density as well as surface effects (for the purpose of our calculations, we assume $\tau \sim 0.5 \times 10^{-9}$ s [36]), $\sigma_{Si} = 14.5 \times 10^{-22}$ m$^2$ is the free-carrier cross section, and $\hbar\omega_0$ is the photon energy [27]. The field profiles obtained through the FDTD method are used to calculate the TPA and FCA effective mode volumes and geometric coefficients in Eqs. (1) and (2) for the first resonant mode ($V_{FCA-M1}=3.28(\lambda_1/n_{si})^3$, $\Gamma_{FCA-M1}=0.998$, $V_{TPA-M1}=4.63(\lambda_1/n_{si})^3$, $\Gamma_{TPA-M1}=0.992$).